\documentclass[3p,preprint]{elsarticle}
\usepackage{etoolbox}
\usepackage{amsmath,amssymb}
\usepackage{hyperref, mathtools, subfigure}
\usepackage{booktabs}
\usepackage{graphicx}
\usepackage{subfigure}
\journal{Communications in Theoretical Physics}

\usepackage{bm}
\usepackage{multirow}
\usepackage{color}
\usepackage{soul,tabularx}
\DeclareUnicodeCharacter{2212}{-}
\DeclareUnicodeCharacter{2009}{-}
\DeclareUnicodeCharacter{03BA}{-}

\begin{document}

\begin{frontmatter}

\title{Theoretical Study of the Photo-stimulated Radio-electric Effect in Asymmetric Semi-parabolic Quantum Wells}
\author[1]{Nguyen Thu Huong}
\author[2]{Nguyen Quang Bau \corref{cor1}}
\author[2]{Cao Thi Vi Ba}
\author[2]{Bui Thi Dung}
\author[2]{Nguyen Cong Toan}
\author[2]{Anh-Tuan Tran \corref{cor2}}
\cortext[cor1]{nguyenquangbau54@gmail.com, nguyenquangbau@hus.edu.vn}
\cortext[cor2]{trananhtuan\_sdh21@hus.edu.vn}
\address[1]{Faculty of Basic Science, Air Defence-Air Force Academy, Kim Son, Son Tay, Hanoi, Vietnam}
\address[2]{Department of Theoretical Physics, University of Science, Vietnam National University, Hanoi, Address: $\rm{No}$ 334 Nguyen Trai, Thanh Xuan, Hanoi, Vietnam}
\begin{abstract}
In this study, based on the quantum kinetic equation approach, we systematically present the radio-electric effect in asymmetric semi-parabolic quantum wells under the influence of a laser radiation field taking into account the electron-longitudinal optical phonon scattering mechanism. The numerical results show that the blue-shift of the maximum peaks in the photon energy range is less than 60 meV. The height of maximum peaks increases according to an exponential rule, depending nonlinearly on the structural parameters of the asymmetric semi-parabolic quantum wells. In the photon energy range greater than 100 meV, the saturated radio-electric field increases with temperature and geometric parameters of the quantum well. The results show the differences between symmetric and asymmetric semi-parabolic quantum wells, highlighting the influence of asymmetric structures on radio-electric effects in two-dimensional quantum well systems.
\end{abstract}
\end{frontmatter}

\section{Introduction}
Quantum wells are thin layers that confine carriers in one dimension, developed by R. Esaki and R. Tsu \cite{f} since the 70s of the 20th century. However, in recent years, problems on optical, electrical, and thermal properties in two-dimensional semiconductor systems, typically quantum wells, have still attracted the attention of many scientists, both theoretically and experimentally. The motivation behind this interest stems from the fact that these systems exhibit significant quantum confinement, resulting in large dipole strengths and huge non-linearities in the transitions between intersubband levels. Consequently, they are considered promising candidates for various applications in photovoltaic \cite{photo, exp1} and thermoelectric materials \cite{thermo}. 

Theoretically, the influence of electronic confinement on kinetic effects in quantum wells has been investigated in detail using many different models and calculation methods. In a recent report on nonlinear absorption of strong electromagnetic waves in quantum wells, C. T. V. Ba \textit{et al.} \cite{p1} demonstrated that the position of the resonance peaks depends not only on the Landau levels that appear due to the influence of the magnetic field and the energy of the phonons but also on the electric sub-band levels that appear due to electron confinement. In this model, the confinement potential is combined by a semi-parabola and a semi-inverse square, which is extended from the semi-parabolic potential model proposed by H. V. Phuc \textit{et al.} \cite{appa, bs1} when they solved the problem of calculating the absorption power of electromagnetic waves using perturbation theory. Using the compact density matrix approach, H. Hassanabadi \textit{et al.} \cite{p2,p3} gave analytical expressions of the second and third-harmonic generation susceptibilities and showed that they are not only strongly influenced by the confinement frequency $\omega_z$ but also by the structural parameter $\beta_z$ of quantum wells. Furthermore, M.K. Bahar \cite{mtc} investigated optical properties of an asymmetric triple $\text{AlGaAs/GaAs}$ quantum well in which the bottom of the potential wells shows an inverse parabolic profile by using the outcome of the iterative solution for the dielectric susceptibility within the compact density matrix formalism. 
The above studies have shown the clear influence of the structural parameters of the confinement potential profile on nonlinear effects in the presence of external electromagnetic waves. However, the photon-drag effect, also known as the radio-electric effect in two-dimensional electron gas (2DEG) systems, has not received much theoretical research attention recently.

The radio-electric phenomenon occurs when a 2DEG system is placed in an electromagnetic wave with a linearly polarized electric field (LPF). Under the effect of light pressure, the electron absorbs a photon of the LPF with suitable energy to transition between subbands. The photon's momentum will force the electron to drift in the direction of the light propagation, causing a photon-drag current in close-circuit conditions \cite{re4, re7,re8} or a radio-electric field (REF) in open-circuit conditions \cite{re1, re2, re3, re5}. Since the 70s of the 20th century, A. F. Gibson \textit{et al.} \cite{rein2} developed from the theory proposed by A. A. Grinberg \cite{rein3} to explain the laser wavelength dependence on the photon drag effect and clarify the contribution of phonon-assisted transitions in the p-type germanium. G. Ribakovs \textit{et al.} \cite{rein1} developed a theoretical model to calculate the longitudinal induced photon drag voltage in tellurium using the Boltzmann transport equation. In 1982, G. M. Shmelev \textit{et al.} \cite{re2} investigated radio-electric phenomena in bulk semiconductors under the influence of a laser radiation field (LRF), in both electron-acoustic phonon and electron-impurity scattering mechanisms using the quantum kinetic equation. Although the above results are very successful in explaining the photon absorption properties and radio-electric current formation in bulk semiconductors, a complete and systematic theoretical study of the radio-electric effect in low-dimensional semiconductors, specifically two-dimensional systems, is still lacking. 

In this paper, we will clarify the influence of the structural parameters of the asymmetric semi-parabolic quantum well and the laser radiation field on the radio-electric effect with the electron-longitudinal optical phonon scattering mechanism using the quantum kinetic equation method. This method was successfully used in G. M. Shmelev's research on kinetic effects in bulk semiconductors, including radio-electric effects \cite{re2} in the 70s of the last century. In recent years, the radio-electric effect has been studied using this method in 2DEG systems such as parabolic quantum wells \cite{re9} and doped superlattices \cite{re3} as well as in other low-dimensional systems such as rectangular quantum wires with an infinite potential \cite{re10}. Inspired by the above works, here, we investigate the radio-electric effect in asymmetric semi-parabolic quantum well under the influence of a laser radiation field on the basis of building a quantum kinetic equation for the non-equilibrium electron distribution function, then calculate the total current density and derive the analytical expression of the radio-electric field under open circuit conditions. The purpose of this study is to quantitatively understand the influence of the geometric and confinement structural parameters of the asymmetric semi-parabolic potential profile of a quantum well on the radio-electric field value as well as the absorption/emission spectrum of electrons in the presence of a laser radiation field. We also present simple new expressions that describe the variation of maximum peak heights, and saturated radio-electric field  with the structural parameters and confinement frequency of the quantum well as well as as external field parameters such as temperature and laser radiation field intensity. The remaining sections of the paper are organized as follows. In Sec. 2, we present the theoretical framework, which introduces the asymmetric semi-parabolic quantum well model with the wave function and energy spectrum of the electron, then solves the quantum kinetic equation and establishes the analytical expression of the radio-electric field. Numerical calculation results and discussion are provided in Sec. 3, and finally, our conclusions are given in Sec. 4. In addition, we also add two appendices including analytic proofs for the form factor of electrons, overlap integral and the symmetric and asymmetric electron distribution functions used in our analytic calculations in Sec. 2. 

\section{Theoretical Framework}
\subsection{Asymmetric confinement potential model of a quantum well}

\begin{figure}[!htb]
\centering
\subfigure[][Asymmetric confinement potential (ACP) and symmetric confinement potential (SCP) of semiconductor quantum well model with different confinement frequencies $\omega_z$ and geometric structural parameters $\beta_z$. \label{1a}]
  {\includegraphics[scale=0.61]{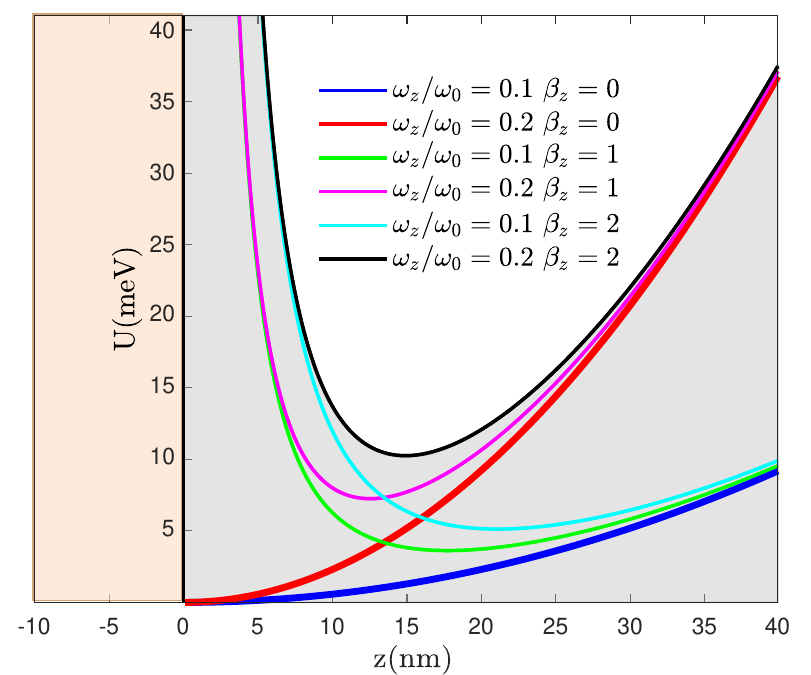}}
\subfigure[][The potential profiles and the associated probability densities ${\left| {{\phi _{\text{n}}}} \left(z\right)\right|^2}$ are displayed for two distinct values of $\beta_z$. Here, ${\omega _z} = 0.2{\omega _0}$.  \label{1b}]
  {\includegraphics[scale=0.64]{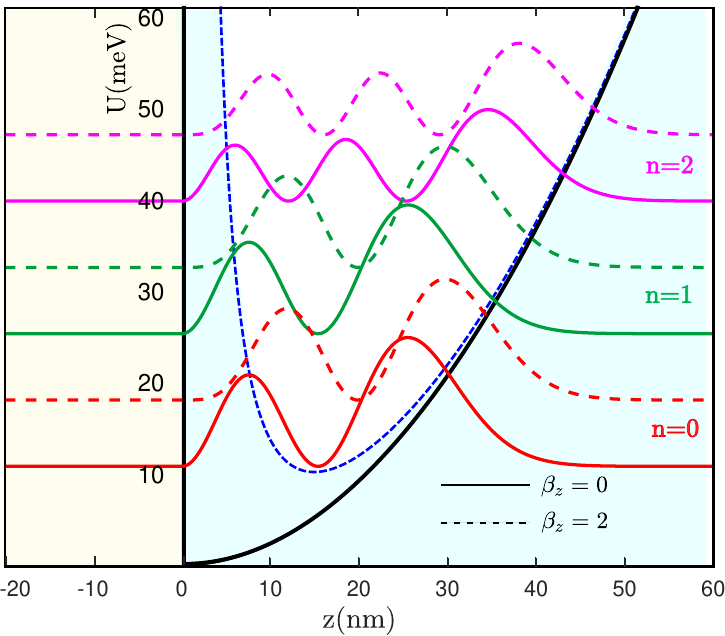}}
\caption{Schematic illustration of the quantum well model with symmetric and asymmetric confinement potentials. }
\label{fig1}
\end{figure}

In this work, we consider the asymmetric confinement potential (ACP) in a quantum well model or so-called infinite semi-parabolic plus semi-inverse squared quantum wells, which is given by \cite{p1, p2} 
\begin{align}\label{uz}
    U\left( z \right) = \left\{ {\begin{array}{*{20}{c}}
  \infty &{z < 0} \\ 
  {\dfrac{1}{2}{m_e}\omega _z^2{z^2} + \dfrac{{{\hbar ^2}{\beta _z}}}{{2{m_e}{z^2}}}}&{z \geqslant 0} 
\end{array}} \right.,
\end{align}
in which, ${m_e} = 0.067{m_0}$ and $\hbar$ are the electron effective mass in the GaAs semiconductor and reduced Planck's constant, respectively, $m_0$ being the free electron rest mass. The z axis is considered the growth direction of the quantum well. The GaAs semiconductor layer is placed between two GaAlAs layers, the first layer is placed in the $z<0$ region, the second layer is not drawn in Fig. \ref{fig1} due to the assumption that the width of the GaAs semiconductor layer is very large in the infinite potential model \cite{f}. Also in Fig. \ref{1a}, the parameters $\omega_z$ and $\beta_z$ are the characteristic parameters for asymmetric semi-parabolic quantum wells. Specifically, $\omega_z$ is the confinement frequency of the quantum well. When $\omega_z$ is larger, the potential profile becomes narrower. In addition, the quantum well is characterised by the geometric structural parameter $\beta_z$. As the value of $\beta_z$ increases, the well gets more distorted, and the asymmetry of the potential profile becomes more pronounced. The infinite asymmetric semi-parabolic quantum well model has been used by many physicists in various problems to study magneto-optical absorption effects \cite{p1, bs2}, the second \cite{p3} and third-harmonic \cite{p2} generation susceptibilities. These studies have shown the strong influence of the geometric structure parameters of the confinement potential on the optical, electrical and magnetic properties of the material. 

In this article, we assume that electrons only emit/absorb and scatter with longitudinal optical (LO) phonons at high temperatures. Therefore, LO phonons are considered non-dispersive and LO phonon energy (in GaAs) $\hbar {\omega _{\bf{q}}}=\hbar {\omega _{{\text{LO}}}} \equiv \hbar {\omega _0} = 36.25{\text{meV}}$ \cite{van, f2}. In Fig. 1, we also present potential models with different profiles depending on the ratio of the quantum well's confinement frequency $\omega_z$ and the LO phonon frequency $\omega_0$.  
Within the framework of the effective mass approximation, the single electron Hamiltonian in
a ACP quantum well reads 
\begin{align}
    {\mathcal{H}_e} = \dfrac{{{\hbar ^2}{\mathbf{k}}_ \bot ^2}}{{2{m_e}}} + \dfrac{{{\hbar ^2}{\mathbf{k}}_z^2}}{{2{m_e}}} + U\left( z \right),
\end{align}
where, ${\mathbf{k}} = \left( {{{\mathbf{k}}_ \bot },{{\mathbf{k}}_z}} \right)$ is the electronic wave vector operator. The corresponding eigenfunctions  and eigenenergies are determined as follows \cite{p1, p2} 
\begin{align}\label{wf}
  {\Psi _{{\text{n}},{{\mathbf{k}}_ \bot }}}\left( {\mathbf{r}} \right) &= {\phi _{{{\mathbf{k}}_ \bot }}}\left( {x,y} \right){\phi _{\text{n}}}\left( z \right), \hfill \\
  {\phi _{{{\mathbf{k}}_ \bot }}}\left( {x,y} \right) &= \dfrac{{\exp \left( {i{{\mathbf{k}}_ \bot } \cdot {{\mathbf{r}}_ \bot }} \right)}}{{\sqrt {{L_x}{L_y}} }}, \label{phxy} \hfill \\
  \left| {\text{n}} \right\rangle &= {\phi _{\text{n}}}\left( z \right) = {\mathcal{C}_{\text{n}}} {z^{2\text{s}}}\exp \left( { - \dfrac{{{z^2}}}{{2\ell _z^2}}} \right)\mathcal{L}_{\text{n}}^\alpha \left( {\dfrac{{{z^2}}}{{\ell _z^2}}} \right), \label{phiz}
\end{align}
\begin{align}\label{pnl}
    {\mathcal{E}_{{\text{n}},{{\mathbf{k}}_ \bot }}} = \dfrac{{{\hbar ^2}{\mathbf{k}}_ \bot ^2}}{{2{m_e}}} + {\mathcal{E}_{\text{n}}} = \dfrac{{{\hbar ^2}{\mathbf{k}}_ \bot ^2}}{{2{m_e}}} + \left( {2{\text{n}} + 1 + \dfrac{{\sqrt {1 + 4{\beta _z}} }}{2}} \right)\hbar {\omega _z},
\end{align}
here, ${{\bf{r}}_ \bot } = \left( {x,y} \right)$ . $L_x$, and $L_y$ are the normalized length in the x-, y-direction, respectively. $\mathrm{n}$ is the level quantization number in z-direction. ${{\rm{s}} = \dfrac{1}{4}\left( {1 + \sqrt {1 + 4{\beta _z}} } \right)}$, and ${{\ell _z} = \sqrt {\dfrac{\hbar }{{{m_e}{\omega _z}}}} }$. ${\mathcal{C}_{\text{n}}} = {\left\{ {{{2{\text{n}}!} \mathord{\left/
 {\vphantom {{2{\text{n}}!} {\left[ {\ell _z^{1 + 4{\text{s}}}\Gamma \left( {2{\text{s}} + {\text{n}} + \dfrac{1}{2}} \right)} \right]}}} \right.
 \kern-\nulldelimiterspace} {\left[ {\ell _z^{1 + 4{\text{s}}}\Gamma \left( {2{\text{s}} + {\text{n}} + \dfrac{1}{2}} \right)} \right]}}} \right\}^{{1 \mathord{\left/
 {\vphantom {1 2}} \right.
 \kern-\nulldelimiterspace} 2}}}$ is the wave function normalization coefficient with $\Gamma(x)$ is the Gamma function. ${\cal L}_{\rm{n}}^{\alpha}(x)$ is the associated Laguerre polynomial with $\alpha  = 2{\text{s}} - \dfrac{1}{2}$. It is important to observe that when $\beta_z$ is equal to zero, the solution in Eqs. \eqref{wf}, \eqref{pnl} will return to the symmetric semi-parabolic potential discussed in problem 10, section 1 of Ref. \cite{asw}. The graph representing the steady states corresponding to the subband energy level and the probability density between different potential profiles (ACP and SCP) is given in Fig. \ref{1b}. 

\subsection{The expression of the REF in ASPQW} 
In this paper, we assume that the system is placed under the influence of external fields as follows \cite{re2, re3} 
\begin{enumerate}
    \item A linearly polarized EMW field (LPF) 
    \begin{align}
        {{\mathbf{E}}^{{\text{LPF}}}}\left( {{\mathbf{r}},t} \right) &= {\mathbf{E}}_0^{{\text{LPF}}}\exp \left( { - i\omega t + i{{\mathbf{k}}_{{\text{LPF}}}} \cdot {\mathbf{r}}} \right) + {\left( {{\mathbf{E}}_0^{{\text{LPF}}}} \right)^ * }\exp \left( {i\omega t - i{{\mathbf{k}}_{{\text{LPF}}}} \cdot {\mathbf{r}}} \right), \label{elpf}\\
        {{\mathbf{B}}^{{\text{LPF}}}}\left( {\bf{r}}, t \right) &= \dfrac{{{{\mathbf{n}}_{{\text{LPF}}}}}}{c} \times {{\mathbf{E}}^{{\text{LPF}}}}\left( {\bf{r}}, t \right),
    \end{align}
    here, $\omega$ is the classical frequency  ($\hbar \omega  \ll \overline \varepsilon  $, with $\overline \varepsilon$ is an average carrier energy), $\left| {{\mathbf{E}}_0^{{\text{LPF}}}} \right|$ is the intensity of the LPF.  ${\left( X \right)^ * }$ is the complex conjugate of $X$. ${{\mathbf{k}}_{{\text{LPF}}}} = \dfrac{\omega }{c}{\mathbf{n}}_{\text{LPF}}$ is wave vector with ${\mathbf{n}}_{\text{LPF}}$ is the unit vector in the direction of the first wave propagation, $c$ is the speed of light in vacuum. However, in the dipole approximation, with the assumption that the spatial variation of the field over the dimensions of the two-dimensional semiconductors systems may be negligible \cite{mmlc, re2, mmlc2}, we can neglect the ${{\mathbf{k}}_{{\text{LPF}}}} \cdot {\mathbf{r}}$ term in the arguments of the functions ${{\mathbf{E}}^{{\text{LPF}}}}\left( {{\mathbf{r}},t} \right)$ and ${{\mathbf{B}}^{{\text{LPF}}}}\left( {{\mathbf{r}},t} \right)$. 
    \item A laser radiation field (LRF): ${{\mathbf{E}}^{{\text{LRF}}}}\left( t \right) = {\mathbf{E}}_0^{{\text{LRF}}}\sin \Omega t$, which is considered to be a linearly polarized high-frequency field with the assumption $\Omega \tau  \geqslant 1$ ($\tau$ is the characteristic relaxation time). $\left| {{\mathbf{E}}_0^{{\text{LRF}}}} \right|$ and $\hbar \Omega $ are the intensity and photon energy of the strong LRF, respectively. 
    \item A DC electric field ${{\mathbf{E}}_0}$ appears due to the open-circuit conditions in all directions of the radio-electric effect. 
\end{enumerate}
We choose ${{\mathbf{E}}^{{\text{LPF}}}\left( t \right)}\parallel {\text{Ox}}{\text{, }}{{\mathbf{B}}^{{\text{LPF}}}\left( t \right)}\parallel {\text{Oz}}{\text{,}}$ and ${\mathbf{n}}_{\text{LPF}}\parallel {{\mathbf{E}}^{{\text{LRF}}}\left( t \right)}\parallel {\text{Oy}}$ and show a schematic illustration of the radio-electric effect as shown in Fig. 2. 
\begin{figure}[!htb]
    \centering
    \includegraphics[scale=0.8]{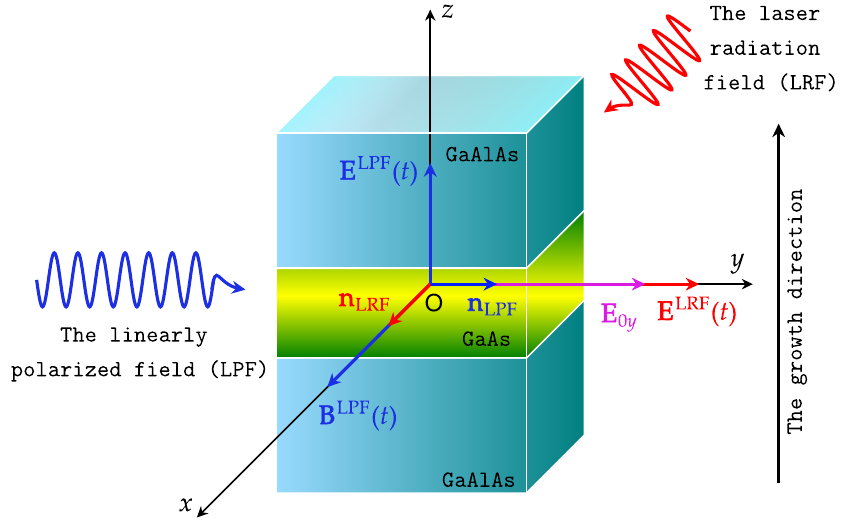}
    \caption{Schematic diagram showing the configuration of the external fields and the sample. }
    \label{fig2}
\end{figure}

Subsequently, we will find the change in the non-equilibrium distribution function of electrons ${\mathcal{F}_{{\text{n}},{{\mathbf{k}}_ \bot }}}$ due to the electron-optical phonon interaction using the quantum kinetic equation \cite{re2, re3}

\begin{align}\label{qke}
\begin{split}
      &\dfrac{{\partial {\mathcal{F}_{{\text{n}},{{\mathbf{k}}_ \bot }}}\left( t \right)}}{{\partial t}} + \left\{ {e{\mathbf{E}}^{\text{LPF}}\left( t \right) + e{{\mathbf{E}}_0} + \dfrac{{e\hbar }}{{{m_e}c}}\left[ {{{\mathbf{k}}_ \bot } \times {\mathbf{B}}^{\text{LPF}}\left( t \right)} \right]} \right\}\dfrac{{\partial {\mathcal{F}_{{\text{n}},{{\mathbf{k}}_ \bot }}}\left( t \right)}}{{\hbar \partial {{\mathbf{k}}_ \bot }}} =  \\
   & \dfrac{{2\pi }}{\hbar }\sum\limits_{{\text{n}},{{\text{n}}^\prime }} {\sum\limits_{\mathbf{q}} {{\mathcal{M}_{{\text{n}},{{\text{n}}^\prime }}}\left( {\mathbf{q}} \right)\sum\limits_{\ell  =  - \infty }^\infty  {\mathcal{J}_\ell ^2\left( {\dfrac{{e{\mathbf{E}^{\text{LRF}}_{0}} \cdot {\mathbf{q}}}}{{{m_e}{\Omega ^2}}}} \right)\left( {{\mathcal{F}_{{{\text{n}}^\prime },{{\mathbf{k}}_ \bot } + {{\mathbf{q}}_ \bot }}}\left( t \right) - {\mathcal{F}_{{\text{n}},{{\mathbf{k}}_ \bot }}}\left( t \right)} \right)\delta \left( {{\mathcal{E}_{{{\text{n}}^\prime },{{\mathbf{k}}_ \bot } + {{\mathbf{q}}_ \bot }}} - {\mathcal{E}_{{\text{n}},{{\mathbf{k}}_ \bot }}} \pm \hbar {\omega _0} - \ell \hbar \Omega } \right)} } },  
\end{split}
\end{align}
here, in the left-hand side of Eq. \ref{qke}, the first term characterizes the rate of change of the non-equilibrium distribution function, the second term characterizes the influence of the LPF on the rate of change of the non-equilibrium distribution function. Meanwhile, the right-hand side of Eq. \ref{qke} is a term that characterizes the influence of the scattering mechanism and the LRF.  ${{{\mathbf{q}}}}$ is the wave vector of phonons, ${\mathcal{J}}_{\ell}\left(x\right)$ is the $\ell-$order Bessel function of real argument $x$, and $\delta \left( x \right)$ being the Dirac Delta function. $ +/- $ sign denotes to the emission/absorption phonon process.
${\mathcal{M}_{{\text{n}},{{\text{n}}^\prime }}}\left( {\mathbf{q}} \right) = {\text{C}}\left( {\mathbf{q}} \right){{\text{J}}_{{\text{n}},{{\text{n}}^\prime }}}\left( {\mathbf{q}} \right)$ is the electron-phonon matrix element determined by the the electron-optical phonon coupling matrix $\text{C}\left(\bf{q}\right)$ in the absence of a radiation field and the electron's form factor ${{\text{J}}_{{\text{n}},{{\text{n}}^\prime }}}\left( {\mathbf{q}} \right)$ characterizes the effect of confinement potential in Eq. \eqref{uz} \cite{p1}
\begin{align}
    {{\left| \text{C}\left( \mathbf{q} \right) \right|}^{2}}&=\frac{2\pi {{e}^{2}}\hbar {{\omega }_{0}}}{{{\varepsilon }_{0}}\kappa {{\left| \mathbf{q} \right|}^{2}}}, \\
    \label{tpj}
    {{\text{J}}_{{\text{n}},{{\text{n}}^\prime }}}\left( {\mathbf{q}} \right) &= \left\langle {\text{n}} \right|\exp \left( { i{q_z}z} \right)\left| {{{\text{n}}^\prime }} \right\rangle {\delta _{{{\mathbf{k}}_ \bot },{{\mathbf{k}}_ \bot } + {{\mathbf{q}}_ \bot }}},
\end{align}
here, $\overline {{\mathcal{N}_{\mathbf{q}}}}  \equiv \overline {{\mathcal{N}_0}}  = {1 \mathord{\left/
 {\vphantom {1 {\left[ {\exp \left( {{{\hbar {\omega _0}} \mathord{\left/
 {\vphantom {{\hbar {\omega _0}} {{k_B}T}}} \right.
 \kern-\nulldelimiterspace} {{k_B}T}}} \right) - 1} \right]}}} \right.
 \kern-\nulldelimiterspace} {\left[ {\exp \left( {{{\hbar {\omega _0}} \mathord{\left/
 {\vphantom {{\hbar {\omega _0}} {{k_B}T}}} \right.
 \kern-\nulldelimiterspace} {{k_B}T}}} \right) - 1} \right]}}$ is the Bose-Einstein distribution function for optical phonons with $k_B$ being the Boltzmann constant. ${{\varepsilon _0}}$ is the permittivity of the vacuum. $\dfrac{1}{\kappa } = \dfrac{1}{{{\kappa _\infty }}} - \dfrac{1}{{{\kappa _0}}}$, with $\kappa _\infty$ and $\kappa _0$ are the static and the high-frequency dielectric constants, respectively. 
${\delta _{m,n}}$ is the Kronecker Delta function. The integral in Eq. \eqref{tpj} is given in \ref{ffe}.

In the linear approximation of the LRF intensity, which corresponds to absorption/emission processes of no more than one photon (i.e., $\ell  = 0, \pm 1$), this limits the terms of the sum to the Bessel function in Eq . \eqref{qke}. We can find the distribution function as \cite{re2} 
\begin{align}
  {\mathcal{F}_{{\text{n}},{{\mathbf{k}}_ \bot }}}\left( t \right) &= {\mathcal{F}_0}\left( {{\mathcal{E}_{{\text{n}},{{\mathbf{k}}_ \bot }}}} \right) - {\mathcal{F}_{\text{p}}}\left( {{\text{n}},{{\mathbf{k}}_ \bot },t} \right), \hfill \label{pb1}\\
  {\mathcal{F}_{\text{p}}}\left( {{\text{n}},{{\mathbf{k}}_ \bot },t} \right) &= \mathcal{F}_{\text{p}}^{(0)}\left( {{\text{n}},{{\mathbf{k}}_ \bot }} \right) + \mathcal{F}_{\text{p}}^{(1)}\left( {{\text{n}},{{\mathbf{k}}_ \bot }} \right)\exp \left( { - i\omega t} \right) + {\left( {\mathcal{F}_{\text{p}}^{(1)}} \right)^ * }\left( {{\text{n}},{{\mathbf{k}}_ \bot }} \right)\exp \left( {i\omega t} \right), \label{pb2}
  \end{align}
where, ${\mathcal{F}_0}\left( {{\mathcal{E}_{{\text{n}},{{\mathbf{k}}_ \bot }}}} \right) = \exp {{\left( {{\mathcal{E}_{\text{F}}} - {\mathcal{E}_{{\text{n}},{{\mathbf{k}}_ \bot }}}} \right)} \mathord{\left/
 {\vphantom {{\left( {{\mathcal{E}_{\text{F}}} - {\mathcal{E}_{{\text{n}},{{\mathbf{k}}_ \bot }}}} \right)} {\left( {{k_B}T} \right)}}} \right.
 \kern-\nulldelimiterspace} {\left( {{k_B}T} \right)}}$ is the Maxwell-Boltzmann distribution function of the non-degenerate  electron gas, in which, ${\mathcal{E}_{\text{F}}}$ being the Fermi energy. $\mathcal{F}_{\text{p}}^{(0)}\left( {{\text{n}},{{\mathbf{k}}_ \bot }} \right)$ and $\mathcal{F}_{\text{p}}^{(1)}\left( {{\text{n}},{{\mathbf{k}}_ \bot }} \right)$ are symmetric and asymmetric parts, respectively. These distribution functions are found as in \ref{hpb}. 

Substituting Eqs. \eqref{pb1} and \eqref{pb2} into the quantum kinetic equation \eqref{qke}, then multiplying $ - \dfrac{{e\hbar }}{{{m_e}}}{{\mathbf{k}}_ \bot }\delta \left( {\mathcal{E} - {\mathcal{E}_{{\text{n}},{{\mathbf{k}}_ \bot }}}} \right)$ into both sides of Eq. \eqref{qke} as in the previous work \cite{baupb, re2}, we transform the quantum kinetic equation for the non-equilibrium distribution function of electrons ${\mathcal{F}_{{\text{n}},{{\mathbf{k}}_ \bot }}}\left( t \right)$ into a new equation for determining the partial current density ${\mathbf{R}}\left( \mathcal{E} \right)$ as follows 

\begin{align}\label{pcd}
    \left[ { - i\omega  + \dfrac{1}{{\tau \left( \mathcal{E} \right)}}} \right]{\mathbf{R}}\left( \mathcal{E} \right) = {\mathbf{Q}} + {\mathbf{S}} + \dfrac{1}{2}\dfrac{{e\hbar }}{{{m_e}c}}\left[ {{{\mathbf{R}}_0}\left( \mathcal{E} \right) \times {\mathbf{B}}^{\text{LPF}}\left( t \right)} \right],
\end{align}
\begin{align}
    \dfrac{{{{\mathbf{R}}_0}\left( \mathcal{E} \right)}}{{\tau \left( \mathcal{E} \right)}} = {{\mathbf{Q}}_0} + {{\mathbf{S}}_0} + \dfrac{1}{2}\dfrac{{e\hbar }}{{{m_e}c}}\left\{ {\left[ {{\mathbf{R}}\left( \mathcal{E} \right) + {{\mathbf{R}}^ * }\left( \mathcal{E} \right)} \right] \times {\mathbf{B}}^{\text{LPF}}\left( t \right)} \right\},
\end{align}

\begin{align}
  {{\mathbf{R}}_0}\left( \mathcal{E} \right) &= \dfrac{{e\hbar }}{{{m_e}c}}\sum\limits_{{\text{n}},{{\mathbf{k}}_ \bot }} {{{\mathbf{k}}_ \bot } \cdot \mathcal{F}_{\text{p}}^{(0)}\left( {{\text{n}},{{\mathbf{k}}_ \bot }} \right)\delta \left( {\mathcal{E} - {\mathcal{E}_{{\text{n}},{{\mathbf{k}}_ \bot }}}} \right)},  \hfill \\
  {\mathbf{R}}\left( \mathcal{E} \right) &= \dfrac{{e\hbar }}{{{m_e}c}}\sum\limits_{{\text{n}},{{\mathbf{k}}_ \bot }} {{{\mathbf{k}}_ \bot } \cdot \mathcal{F}_{\text{p}}^{(1)}\left( {{\text{n}},{{\mathbf{k}}_ \bot }} \right)\delta \left( {\mathcal{E} - {\mathcal{E}_{{\text{n}},{{\mathbf{k}}_ \bot }}}} \right)},  \hfill \\
  {{\mathbf{Q}}_0}\left( \mathcal{E} \right) &=  - \dfrac{{e\hbar }}{{{m_e}c}}\sum\limits_{{\text{n}},{{\mathbf{k}}_ \bot }} {\left[ {e{{\mathbf{E}}_0} \cdot \dfrac{{\partial {\mathcal{F}_0}\left( {{\mathcal{E}_{{\text{n}},{{\mathbf{k}}_ \bot }}}} \right)}}{{\hbar \partial {{\mathbf{k}}_ \bot }}}} \right]{{\mathbf{k}}_ \bot }\delta \left( {\mathcal{E} - {\mathcal{E}_{{\text{n}},{{\mathbf{k}}_ \bot }}}} \right)},  \hfill \\
  {\mathbf{Q}}\left( \mathcal{E} \right) &=  - \dfrac{{e\hbar }}{{{m_e}c}}\sum\limits_{{\text{n}},{{\mathbf{k}}_ \bot }} {\left[ {e{\mathbf{E}}^{\text{LPF}}\left( t \right) \cdot \dfrac{{\partial {\mathcal{F}_0}\left( {{\mathcal{E}_{{\text{n}},{{\mathbf{k}}_ \bot }}}} \right)}}{{\hbar \partial {{\mathbf{k}}_ \bot }}}} \right]{{\mathbf{k}}_ \bot }\delta \left( {\mathcal{E} - {\mathcal{E}_{{\text{n}},{{\mathbf{k}}_ \bot }}}} \right)}  ,
\end{align}
\begin{align}
  {{\mathbf{S}}_0}\left( \mathcal{E} \right) &=  - \dfrac{{2\pi e}}{{{m_e}}}\sum\limits_{{{\text{n}}^\prime },{\text{n}}} {\sum\limits_{{{\mathbf{k}}_ \bot },{\mathbf{q}}} {{{\left| {{\mathcal{M}_{{\text{n}},{{\text{n}}^\prime }}}\left( {\mathbf{q}} \right)} \right|}^2}\overline {{\mathcal{N}_{\mathbf{q}}}} {{\mathbf{k}}_ \bot }\delta \left( {\mathcal{E} - {\mathcal{E}_{{\text{n}},{{\mathbf{k}}_ \bot }}}} \right)\Theta \left( {{\mathbf{q}},{\mathbf{E}}_0^{{\text{LRF}}},\Omega } \right)\left[ {\mathcal{F}_{\text{p}}^{(0)}\left( {{{\text{n}}^\prime },{{\mathbf{k}}_ \bot } + {{\mathbf{q}}_ \bot }} \right) - \mathcal{F}_{\text{p}}^{(0)}\left( {{\text{n}},{{\mathbf{k}}_ \bot }} \right)} \right]} },  \hfill \\
  {\mathbf{S}}\left( \mathcal{E} \right) &=  - \dfrac{{2\pi e}}{{{m_e}}}\sum\limits_{{{\text{n}}^\prime },{\text{n}}} {\sum\limits_{{{\mathbf{k}}_ \bot },{\mathbf{q}}} {{{\left| {{\mathcal{M}_{{\text{n}},{{\text{n}}^\prime }}}\left( {\mathbf{q}} \right)} \right|}^2}\overline {{\mathcal{N}_{\mathbf{q}}}} {{\mathbf{k}}_ \bot }\delta \left( {\mathcal{E} - {\mathcal{E}_{{\text{n}},{{\mathbf{k}}_ \bot }}}} \right)\Theta \left( {{\mathbf{q}},{\mathbf{E}}_0^{{\text{LRF}}},\Omega } \right)\left[ {\mathcal{F}_{\text{p}}^{(1)}\left( {{{\text{n}}^\prime },{{\mathbf{k}}_ \bot } + {{\mathbf{q}}_ \bot }} \right) - \mathcal{F}_{\text{p}}^{(1)}\left( {{\text{n}},{{\mathbf{k}}_ \bot }} \right)} \right]} }, 
\end{align}
\begin{align}
    \begin{split}
  \Theta \left( {{\mathbf{q}},{\mathbf{E}}_0^{{\text{LRF}}},\Omega } \right) &= \delta \left( {{\mathcal{E}_{{{\text{n}}^\prime },{{\mathbf{k}}_ \bot } + {{\mathbf{q}}_ \bot }}} - {\mathcal{E}_{{\text{n}},{{\mathbf{k}}_ \bot }}} \pm \hbar {\omega _{\mathbf{q}}}} \right) - {\left( {\dfrac{{e{\mathbf{E}}_0^{{\text{LRF}}} \cdot {\mathbf{q}}}}{{2{m_e}{\Omega ^2}}}} \right)^2} \hfill \\
   &\times \left[ {2\delta \left( {{\mathcal{E}_{{{\text{n}}^\prime },{{\mathbf{k}}_ \bot } + {{\mathbf{q}}_ \bot }}} - {\mathcal{E}_{{\text{n}},{{\mathbf{k}}_ \bot }}} \pm \hbar {\omega _{\mathbf{q}}}} \right) - \delta \left( {{\mathcal{E}_{{{\text{n}}^\prime },{{\mathbf{k}}_ \bot } + {{\mathbf{q}}_ \bot }}} - {\mathcal{E}_{{\text{n}},{{\mathbf{k}}_ \bot }}} \pm \hbar {\omega _{\mathbf{q}}} - \hbar \Omega } \right)} \right] ,
    \end{split}
\end{align}
here,  $\tau \left( \mathcal{E} \right)$ is the relaxation time which can be approximated by power laws in the form $\tau \left( \mathcal{E} \right) = {\tau _0}\sqrt {{\mathcal{E} \mathord{\left/
 {\vphantom {\mathcal{E} {{\mathcal{E}_{\text{F}}}}}} \right.
 \kern-\nulldelimiterspace} {{\mathcal{E}_{\text{F}}}}}} $ \cite{re2, lund}, ${\tau _0}$ is the constant relaxation time which appears when solving Eq. \eqref{qke} in the constant relaxation time approximation (see \ref{hpb}). 

From the expression of partial current density in Eq. \eqref{pcd}, we determine the total current density as \cite{re2}
\begin{align}
    {\mathbf{J}} = {\mathbf{J}}\left( t \right) + {{\mathbf{J}}_0} = \int\limits_0^\infty  {\left[ {{\mathbf{R}}\left( \mathcal{E} \right)\exp \left( { - i\omega t} \right) + {{\mathbf{R}}^ * }\left( \mathcal{E} \right)\exp \left( {i\omega t} \right) + {{\mathbf{R}}_0}\left( \mathcal{E} \right)} \right]d\mathcal{E}} .
\end{align}

Finally, with open-circuit electrical
boundary conditions, i.e. $\textbf{J} = \textbf{0}$, and ${\mathbf{n}}\parallel {{\mathbf{E}}^{{\text{LRF}}}}\parallel {\text{Oy}}$ \cite{re2}, we establish the expression for the longitudinal electric-radio field as 
\begin{align}
  {{\text{E}}_{0y}} &= \dfrac{{\omega {\tau _0}}}{{1 + {\omega ^2}{\tau _0^2}}}\dfrac{{\varpi  + \vartheta }}{{\rho  + \varsigma }}\left| {{\mathbf{E}}_0^{{\text{LPF}}}} \right|, \hfill \\
  \varpi  &= \dfrac{{{e^2}{k_B}T}}{{2\pi {\hbar ^2}}}\sum\limits_{\text{n}} {\exp \left( {\dfrac{{{{\mathcal{E}}_{\text{F}}} - {{\mathcal{E}}_{\text{n}}}}}{{2{k_B}T}}} \right)} , \hfill \\
  \vartheta  &=  - \dfrac{{1 - {\omega ^2}{\tau _0^2}}}{{1 + {\omega ^2}{\tau _0^2}}}\sum\limits_{{\text{n}},{{\text{n}}^\prime }} {\mathcal{A}\left( {2{k_B}T - {{\text{D}}_{{\text{n}},{{\text{n}}^\prime }}} - {k_B}T\exp \dfrac{{ \pm \hbar {\omega _0} - \hbar \Omega }}{{{k_B}T}}} \right)} , \hfill \\
  \rho  &= \sum\limits_{{\text{n}},{{\text{n}}^\prime }} {\mathcal{A}\left( {{{\text{D}}_{{\text{n}},{{\text{n}}^\prime }}} + {k_B}T\exp \dfrac{{ \pm \hbar {\omega _0} - \hbar \Omega }}{{{k_B}T}}} \right)} , \hfill \\
  \varsigma  &= \dfrac{{{e^2}{k_B}T}}{{2\pi {\hbar ^2}}}\sum\limits_{\text{n}} {\exp \left( {\dfrac{{{{\mathcal{E}}_{\text{F}}} - {{\mathcal{E}}_{\text{n}}}}}{{{k_B}T}}} \right)}  - \dfrac{{1 + {\omega ^2}\tau _0^2}}{{1 - {\omega ^2}\tau _0^2}}\vartheta , \hfill \\
  \mathcal{A} &= \dfrac{{{e^6}{\omega _0}{\tau _0}{{\left| {{\mathbf{E}}_0^{{\text{LRF}}}} \right|}^2}}}{{16{\varepsilon _0}{m_e}{{\left( {\hbar \omega } \right)}^4}\kappa }}\overline {{\mathcal{N}_0}} \left| {{{\text{G}}_{{\text{n}},{{\text{n}}^\prime }}}} \right|\exp \left( {\dfrac{{{\mathcal{E}_{\text{F}}} - {\mathcal{E}_{\text{n}}}}}{{{k_B}T}}} \right)\exp \left( { - \dfrac{{{{\text{D}}_{{\text{n}},{{\text{n}}^\prime }}}}}{{2{k_B}T}}} \right), \hfill \\
  {{\text{G}}_{{\text{n}},{{\text{n}}^\prime }}} &= \int\limits_{ - \infty }^{ + \infty } {{{\left| {{{\text{J}}_{{\text{n}},{{\text{n}}^\prime }}}\left( {{q_z}} \right)} \right|}^2}d{q_z}}, \label{gnn} \hfill \\
  {{\text{D}}_{{\text{n}},{{\text{n}}^\prime }}} &= {\mathcal{E}_{{{\text{n}}^\prime }}} - {\mathcal{E}_{\text{n}}} \pm \hbar {\omega _0} - \hbar \Omega.
\end{align}

The integral in Eq. \eqref{gnn} is called the overlap integral associated with the transition between the $\left| {\text{n}} \right\rangle $ state and $\left| {{{\text{n}}^\prime }} \right\rangle $ state. This overlap integral is calculated in detail in the \ref{ffe}.  

The analytical expression of the REF obtained here is extremely complicated, depending on many parameters of the external fields as well as the characteristic parameters of the asymmetric semi-parabolic quantum well such as the temperature, the intensities and photon energies of LRF and LPF, the confinement frequency and geometric structural parameter $\beta_z$ of quantum wells. In the next section, we will provide detailed numerical calculations to elucidate these effects. Furthermore, also with the help of computer programs, we offer a comprehensive analysis of the physical implications of the analytical findings derived in this work.

\section{Results and Discussions}
In this section, to clarify the difference between symmetric and asymmetric semi-parabolic quantum wells as well as the influence of the external fields on the REF under different temperature conditions and structural parameters of the quantum well models, we present detailed numerical calculations for the analytical results obtained in the previous section. The parameters used for calculations for the ${\text{GaAs}}/{\text{GaAlAs}}$ heterostructures are given in the table below

\begin{table}[!htb]
\centering
\caption{Main input parameters of our calculation model.}
\begin{tabular}{clcc}
\toprule
\hline
\multicolumn{1}{c}{Symbol} & \multicolumn{1}{c}{Definition} & Value & Unit \\ \hline
$\hbar {\omega _0}$ & LO phonon energy \cite{van, f2}  & 36.25 & meV \\
$\tau_0$ & Constant relaxation time \cite{rta} & $1.0 \times {10^{ - 12}}$ & s \\
\begin{tabular}[c]{@{}l@{}}${{{m_e}} \mathord{\left/
 {\vphantom {{{m_e}} {{m_0}}}} \right.
 \kern-\nulldelimiterspace} {{m_0}}}$\end{tabular} & \begin{tabular}[c]{@{}l@{}}Ratio of the effective mass \\ to the rest mass of electrons \cite{p1,f2,ts}\end{tabular} & 0.067 &  \\
$\hbar\omega$ & Photon energy of LPF & $50$ & ${\text{meV}}$ \\
$\left| {{\mathbf{E}}_0^{{\text{LPF}}}} \right|$ & Intensity of LPF & ${\text{5}} \times {\text{1}}{{\text{0}}^{4}}$ & V/cm \\
${\varepsilon _0}$ & Vacuum permittivity & $8.85 \times {10^{ - 12}}$ & ${\text{F/m}}$ \\
${\kappa _0}$ & Static dielectric constant \cite{ts} & 12.53 &  \\
${\kappa _\infty }$ & Optical dielectric constant \cite{ts} & 10.82 &  \\
${\mathcal{E}_{\text{F}}}$ & Fermi energy \cite{p1}& $50$ & ${\text{meV}}$ \\ 
\hline
\bottomrule
\end{tabular}
\label{tab1}
\end{table}

In Fig. \ref{3a}, we show the dependence of the REF on the LRF photon energy under the influence of temperature. The results are evaluated in both cases of quantum wells with symmetric and asymmetric profiles. We can see that temperature significantly affects the value of the REF but does not change the position of the resonance peak. This can be explained by the electron-phonon resonance condition. This can be explained by the electron-phonon resonance condition. In the approximation when studying the radio-electric effect, we only consider the absorption/emission of one photon by the LRF. Therefore, we have the electron-phonon resonance condition ${\mathcal{E}_{{{\text{n}}^\prime }}} - {\mathcal{E}_{\text{n}}} \pm \hbar {\omega _0} \pm \hbar \Omega  = 0$, which is a direct consequence of the energy-momentum conservation law or also called the selection rules, allowing us to precisely determine the position of the maximum peaks. To simplify and easily clarify the meaning of each maximum peak, we only consider the transition process between two states: $\left| 0 \right\rangle $  and $\left| 1 \right\rangle $. In Fig. \ref{3a}, we estimate the positions of the three maximum peaks corresponding to the quantum well's confinement frequency ${\omega _z} = 0.1{\omega _0}$. First, the first absorption peak appears at $\hbar {\Omega _1} = {\mathcal{E}_{\text{1}}} - {\mathcal{E}_0} = 7.3{\text{meV}}$. This implies that the electron absorbs a photon of the LRF to transition from $\left| 0 \right\rangle $ state to $\left| 1 \right\rangle $ state without accompanying absorption or phonon emission. The second maximum peak appears at $\hbar {\Omega _2} = {\mathcal{E}_0} - {\mathcal{E}_1} + \hbar {\omega _0} = 29.2 {\text{meV}}$, showing that the electron absorbs one phonon and emits one photon to transition from the first excited state back to the ground state. Finally, the third maximum peak appears at $\hbar {\Omega _3} = {\mathcal{E}_1} - {\mathcal{E}_0} - \hbar {\omega _0} = 43.8{\text{meV}}$. This peak is the result of the transition from the ground state to the first excited state of the electron via the one-photon absorption process accompanied by the one-phonon emission process. A third absorption peak position was found in a recent study in which the authors did not consider phonon absorption. The third absorption peak position has been mentioned in a recent study (not considering phonon absorption) \cite{p1}. In addition, we can also see that the value of the structural parameter $\beta_z$ strongly affects the value of the REF as well as the height of the maximum peaks. In other words, the REF in an asymmetric semi-parabolic quantum well has a significantly larger value than in the case of a symmetric semi-parabolic quantum well. This can be briefly explained as follows. In Fig. \ref{1a}, it can be seen that the potential profile becomes smaller as $\beta_z$ increases, leading to stronger electron confinement. This means that the scattering probability between electrons, phonons and photons increases and leads to an increase in the REF. We will discuss this in more detail in a quantitative calculation later. 

From the above resonance condition and Fig. \ref{3a}, it can be seen that the value of $\beta_z$ does not affect the position of the maximum peaks. This is also consistent with the numerical calculations in Fig. \ref{fig4}. In Fig. \ref{fig4}, and Fig. \ref{fig5}, we also add quantitative calculations to clarify the dependence of the peak positions on the confinement frequency of the quantum well. The results in Fig. \ref{fig5} can also be explained from resonance conditions. Indeed, considering the transitions between the ground state and the first excited state, the position of the maximum peaks is determined as follows   
\begin{align}\label{hOhwz}
    \hbar {\Omega _1} = 2\hbar {\omega _z};\hbar {\Omega _2} = \hbar {\omega _0} - 2\hbar {\omega _z}; \text{and }\hbar {\Omega _3} = \hbar {\omega _0} + 2\hbar {\omega _z}.
\end{align}
It can be seen that the structural parameter $\beta_z$ does not appear in Eq. \eqref{hOhwz}, so the positions of the maximum peaks do not depend on $\beta_z$. This means that there is no shift when $\beta_z$ changes. This conclusion is consistent with another previous study using the compact-density-matrix method \cite{p3}. 
Also from Eq. \eqref{hOhwz}, we see a linear dependence on $\omega_z$ of the positions of the maximum peaks, in which the positions of peaks (1), and (3) increase with $\omega_z$ while the position of peak (2) decreases with $\omega_z$. The rightward shift toward the larger photon energy domain of the maximum peaks (1) and (3) due to one-photon absorption process is called blue-shift, which has been reported in previous studies on parabolic quantum wells \cite{bs1}, symmetric \cite{appa} and asymmetric \cite{p1,bs2} semi-parabolic quantum wells. Another interesting thing to note here is that if the 2DEG system under consideration is placed in a perpendicular uniform magnetic field, the magnetic field will create a confinement potential similar in shape to the parabolic potential. Formally, with the substitution of the confinement frequency $\omega_z$ to the cyclotron frequency $\omega_c$ (Landau levels appear under resonance conditions), the above-mentioned blue-shift will also appear when $\omega_c$ is increased by increasing the magnetic field strength. This is not only true for 2DEG systems such as quantum wells, quantum rings \cite{qtr, qtr2} or compositional superlattices \cite{cs}  but also true for other low-dimensional systems such as quantum dots \cite{qdhp,qd2}, and quantum wires \cite{qwr}. In addition to the geometric structure parameter $\beta_z$ of the semi-parabolic quantum well and the temperature, another parameter, the intensity of the laser radiation field, also does not affect the position of the resonance peaks. Fig. \ref{3b} shows the influence of laser radiation field intensity on the spectral lines. When the laser radiation field is absent, the value of REF remains constant at about 27.8 V/cm and no resonance peaks appear. However, in the presence of a laser radiation field, an absorption/emission spectrum appears due to the electron transitions discussed above. A notable point here is that in the high photon energy region, the REF value gradually decreases to a saturated value. Besides, the saturated value of REF also depends on the temperature and structural parameters of the quantum well, which will be discussed later.

Now, we turn our attention to the height of maximum peaks (HMP) in Fig. \ref{fig5}. As discussed above, as the confinement frequency increases, the REF value increases and naturally the HMP also increases go up. We present calculations that quantify the increase of HMP with both the confinement frequency and the structural parameters of the quantum well simultaneously in Figs. \ref{fig6}, and \ref{fig7}. In Fig. \ref{6a}, we present the dependence of the HMP of each maximum peak on the ratio ${{{\omega _z}} \mathord{\left/
 {\vphantom {{{\omega _z}} {{\omega _0}}}} \right.
 \kern-\nulldelimiterspace} {{\omega _0}}}$ at different values of $\beta_z$, with the interpolation exponential function found by numerical analysis of the form 
${\text{HM}}{{\text{P}}_i} = {\mathcal{G}_i} \times \exp \left( {{\mathcal{K}_i} \times \dfrac{{{\omega _z}}}{{{\omega _0}}}} \right)$. Similarly, in Fig. \ref{6b}, we present the HMP of each peak as a function of the ratio $\beta_z$ at different values of ${{{\omega _z}} \mathord{\left/
 {\vphantom {{{\omega _z}} {{\omega _0}}}} \right.
 \kern-\nulldelimiterspace} {{\omega _0}}}$, with the same interpolation rule ${\text{HM}}{{\text{P}}_i} = {\mathcal{T}_i} \times \exp \left( {{\mathcal{Q}_i} \times {\beta _z}} \right)$, with $i=1, 2,$ and 3, corresponding to the first, second, and third maximum peaks. The interpolation parameters ${\mathcal{G}_i}$, ${\mathcal{K}_i}$, ${\mathcal{T}_i}$, and ${\mathcal{Q}_i}$ are given in Tabs. \ref{tab2}, and \ref{tab3} below

\begin{table}[!htb]
\centering
\caption{The curve interpolation parameters show the dependence of HMP on the ratio ${{{\omega _z}} \mathord{\left/
 {\vphantom {{{\omega _z}} {{\omega _0}}}} \right.
 \kern-\nulldelimiterspace} {{\omega _0}}}$ of HMP in Fig. \ref{6a}.}
\label{tab2}
\begin{tabular}{ccccc}
\toprule
\hline
\multicolumn{5}{c}{${\text{HM}}{{\text{P}}_i} = {\mathcal{G}_i} \times \exp \left( {{\mathcal{K}_i} \times \dfrac{{{\omega _z}}}{{{\omega _0}}}} \right)$} \\ \hline
\multirow{2}{*}{\begin{tabular}[c]{@{}c@{}}Position of\\ maximum peak $\hbar \Omega_i$  \\ \end{tabular}} & \multicolumn{2}{c}{$\beta_z=0$} & \multicolumn{2}{c}{$\beta_z=2$} \\ \cline{2-5} 
 & ${\mathcal{G}_i}$ (V/cm) & ${\mathcal{K}_i}$ & ${\mathcal{G}_i}$ (V/cm) & ${\mathcal{K}_i}$ \\ \hline
i = 1 & 10.82 & 7.176 & 10.73 & 10.04 \\ \hline
i = 2 & 9.919 & 7.534 & 10.83 & 10.4 \\ \hline
i = 3 & 10.08 & 6.413 & 10.01 & 9.26 \\ \hline
\bottomrule
\end{tabular}
\end{table}
 
\begin{table}[!htb]
\centering
\caption{The curve interpolation parameters show the dependence of HMP on the structural parameter $\beta_z$ in Fig. \ref{6b}.}
\label{tab3}
\begin{tabular}{ccccc}
\toprule
\hline
\multicolumn{5}{c}{${\text{HM}}{{\text{P}}_i} = {\mathcal{T}_i} \times \exp \left( {{\mathcal{Q}_i} \times {\beta _z}} \right)$} \\ \hline
\multirow{2}{*}{\begin{tabular}[c]{@{}c@{}}Position of\\ maximum peak $\hbar \Omega_i$ \\ \end{tabular}} & \multicolumn{2}{c}{\begin{tabular}[c]{@{}c@{}}${{{\omega _z}} \mathord{\left/  {\vphantom {{{\omega _z}} {{\omega _0}}}} \right.  \kern-\nulldelimiterspace} {{\omega _0}}} = 0.1$\end{tabular}} & \multicolumn{2}{c}{\begin{tabular}[c]{@{}c@{}}${{{\omega _z}} \mathord{\left/ {\vphantom {{{\omega _z}} {{\omega _0}}}} \right. \kern-\nulldelimiterspace} {{\omega _0}}} = 0.2$\end{tabular}} \\ \cline{2-5} 
 & ${\mathcal{T}_i}$ (V/cm) & ${\mathcal{Q}_i}$ & ${\mathcal{T}_i}$ (V/cm) & ${\mathcal{Q}_i}$ \\ \hline
i = 1 & 23.32 & 0.1326 & 48.97 & 0.2605 \\ \hline
i = 2 & 22.14 & 0.1325 & 48.16 & 0.2605 \\ \hline
i = 3 & 20.03 & 0.1322 & 39.02 & 0.2599 \\ \hline
\bottomrule
\end{tabular}
\end{table}

From Tab. \ref{tab2}, we see that the ${\mathcal{G}}_i$ value is not much different between the maximum peaks, and also not much different when changing the structural parameter $\beta_z$ of the quantum well. Meanwhile, in Tab. \ref{tab3}, when $\omega_z$ doubles, the ${\mathcal{Q}}_i$ value also doubles. Notwithstanding, a simultaneous dependence on the structural parameters of the quantum well of HMPs becomes much more complicated. Fig. \ref{fig7} shows us a simultaneous monotonic increase in both $\omega_z$ and $\beta_z$. This proves that the asymmetric quantum well model has a clearer and stronger influence on the optical and electrical properties, including radio-electric effects, than the symmetric quantum well model. To the best of our knowledge at the present time, there have not been any experimental observations to verify these laws. Therefore, we hope that the above new findings can play a predictive role and guide future experiments. 

To better understand the influence of LRF's photon energy on the radio-electric effect, in Fig. \ref{fig8}, we  depict the variation of the REF with photon energy for different values of the temperature, mainly focusing on the energy region higher than 100 meV. It is easy to observe that besides the three maximum peaks appearing at the positions calculated above, the value of REF decreases rapidly to a saturated value under different temperature conditions in the photon energy range higher than 100 meV. In general, the higher the temperature, the larger the value of saturated REF. The value of saturated REF in the case of asymmetric potential profile is larger than the value of saturated REF in the case of symmetric potential profile. To have a more detailed view, we quantitatively investigate the temperature and magnetic field variation of the contour plot of the saturated REF, as shown in Fig. \ref{9a}. It can be seen that, in the case of a symmetric semi-parabolic quantum well ($\beta_z = 0$), the saturated REF increases more slowly with temperature than in the case of an asymmetric semi-parabolic well (${\beta _z} \ne 0$). Quantitatively, we illustrate the temperature-dependent saturated REF for different values of the confinement frequency and structural parameters of the quantum well given in Fig. \ref{9b} with the rules 
\begin{align}\label{stet}
    {\text{SREF}} = a\exp \left( {\dfrac{b}{T}} \right) + c\sqrt T ,
\end{align}
and interpolation parameters in Tab. \ref{tab4}. We can see that this expression fits the saturated REF data (filled triangles, diamonds, squares, and circles) very well. Unfortunately, there is currently a lack of empirical evidence to substantiate our forecast. We expect that our findings will serve as a valuable guide for future experimental research. 

\begin{table}[!htb]
\centering
\caption{Law of temperature dependence of saturated REF (SREF) and fitted results of parameters in Eq. \eqref{stet}.}
\label{tab4}
\begin{tabular}{lcccc}
\toprule
\hline
\multicolumn{1}{c}{Interpolation rules} & \multicolumn{4}{c}{${\text{SREF}} = a\exp \left( {{b \mathord{\left/{\vphantom {b T}} \right.\kern-\nulldelimiterspace} T}} \right) + c\sqrt T $} \\ \hline
\multicolumn{1}{c}{\multirow{2}{*}{Fitted parameters}} & \multicolumn{2}{c}{$\beta_z=0$} & \multicolumn{2}{c}{$\beta_z=2$} \\ \cline{2-5} 
\multicolumn{1}{c}{} & ${{{\omega _z}} \mathord{\left/ {\vphantom {{{\omega _z}} {{\omega _0}}}} \right. \kern-\nulldelimiterspace} {{\omega _0}}} = 0.1$ & ${{{\omega _z}} \mathord{\left/ {\vphantom {{{\omega _z}} {{\omega _0}}}} \right. \kern-\nulldelimiterspace} {{\omega _0}}} = 0.2$ & ${{{\omega _z}} \mathord{\left/ {\vphantom {{{\omega _z}} {{\omega _0}}}} \right. \kern-\nulldelimiterspace} {{\omega _0}}} = 0.1$ & ${{{\omega _z}} \mathord{\left/ {\vphantom {{{\omega _z}} {{\omega _0}}}} \right. \kern-\nulldelimiterspace} {{\omega _0}}} = 0.2$ \\ \hline
$a$ (V/cm) & 129.3 & 263.6 & 179.4 & 501.6 \\ \hline
$b$ (K) & $-114.8$ & $-116.6$ & $-103.3$ & $-94.4$ \\ \hline
$c \left( {{{\text{V}} \mathord{\left/{\vphantom {{\text{V}} {\left( {{\text{cm}}{\text{.}}{{\text{K}}^{{1 \mathord{\left/{\vphantom {1 2}} \right.\kern-\nulldelimiterspace} 2}}}} \right)}}} \right.\kern-\nulldelimiterspace} {\left( {{\text{cm}}{\text{.}}{{\text{K}}^{{1 \mathord{\left/{\vphantom {1 2}} \right.\kern-\nulldelimiterspace} 2}}}} \right)}}} \right)$ & $-3.8$ & $-7.6$ & $-5.93$ & $-17.72$ \\ \hline
\bottomrule
\end{tabular}
\end{table}

\section{Conclusions}
In this paper, we have presented a systematic theoretical study of radio-electric effects in two-dimensional quantum wells with symmetric and asymmetric semi-parabolic potential profiles using the quantum kinetic equation method. The analytical results include the influence of laser radiation field and electron-phonon interaction in the high temperature region. The temperature and structural parameters as well as the confinement frequency of the asymmetric semi-parabolic quantum well have a strong and nonlinear influence on the radio-electric field appearing along the direction of the incident light. Besides, we also observed the appearance of resonant peaks in the photon energy region lower than 80 meV following the energy-momentum conservation law. Our numerical results show that when increasing the confinement frequency, a blue shift of the resonance peaks appears similar to that in previous studies. The height of the maximum peaks increases with the parameters of the quantum well with the exponential rules predicted in Tabs. \ref{tab2} and \ref{tab3}. Furthermore, in the photon energy range greater than 100 meV, the appearance of a saturated REF depends nonlinearly on temperature. Nevertheless, there are still no direct experimental observations on the asymmetric semi-parabolic quantum well model to verify our predictions about the HMP, and saturated REF. We hope that our theoretical results will be a good orientation for future experiments.


\section*{Acknowledgement}
This research is financially supported by Vietnam National University, Hanoi - Grant number QG. 23.06. 
\appendix 
\section{The form factor and overlap integral of electrons}\label{ffe}
In this Appendix, we briefly introduce how to determine the form factor of electron (in Eq. \eqref{tpj}) in two-dimensional quantum wells similar to the method presented in Refs. \cite{exp1, mt1}. 
First of all, the expression for the electronic form factor is given by \cite{exp1,mt1,mt2} 
\begin{align}\label{jnn}
    \begin{split}
  {{\text{J}}_{{\text{n}},{{\text{n}}^\prime }}}\left( {\mathbf{q}} \right) &= \left\langle {{\Psi _{{\text{n}},{{\mathbf{k}}_ \bot }}}\left( {\mathbf{r}} \right)} \right|\exp \left( {i{\mathbf{q}} \cdot {\mathbf{r}}} \right)\left| {{\Psi _{{{\text{n}}^{\prime}},{{\mathbf{k}}_ \bot }}}\left( {\mathbf{r}} \right)} \right\rangle  \hfill \\
   &= \left\langle {{\phi _{{{\mathbf{k}}_ \bot }}}\left( {x,y} \right)} \right|\exp \left[ {i\left( {{q_x}x + {q_y}y} \right)} \right]\left| {{\phi _{{{\mathbf{k}}_ \bot }}}\left( {x,y} \right)} \right\rangle \left\langle {{\phi _{\text{n}}}\left( z \right)} \right|\exp \left( {i{q_z}z} \right)\left| {{\phi _{\text{n}}}\left( z \right)} \right\rangle  \equiv {{\text{J}}_1} \cdot {{\text{J}}_2}, 
    \end{split}
\end{align}

From the expressions of the wave functions ${\phi _{{{\mathbf{k}}_ \bot }}}\left( {x,y} \right)$ and ${\phi _{\text{n}}}\left( z \right)$ in Eqs. \eqref{phxy}, \eqref{phiz}, we obtain
\begin{align}\label{j1}
\begin{split}
      {{\text{J}}_1} &= \langle {\phi _{{{\mathbf{k}}_ \bot }}}\left( {x,y} \right)|\exp \left[ { - i\left( {{q_x}x + {q_y}y} \right)} \right]\left| {{\phi _{{\mathbf{k}}_ \bot ^\prime }}\left( {x,y} \right)} \right\rangle  \hfill \\
   &= \left\{ {\dfrac{1}{{{L_x}}}\int\limits_0^{{L_x}} {\exp \left[ { - ix\left( {{k_x} + {q_x} - k_x^\prime } \right)} \right]dx} } \right\}\left\{ {\dfrac{1}{{{L_y}}}\int\limits_0^{{L_y}} {\exp \left[ { - iy\left( {{k_y} + {q_y} - k_y^\prime } \right)} \right]dy} } \right\} \hfill \\
   &= {\delta _{{k_x} + {q_x},k_x^\prime }}{\delta _{{k_y} + {q_y},k_y^\prime }} \equiv {\delta _{{{\mathbf{k}}_ \bot } + {{\mathbf{q}}_ \bot },{{\mathbf{k}}_ \bot }}}, 
\end{split}
\end{align}
\begin{align}\label{j2}
    {{\text{J}}_2} = \langle {\phi _{\text{n}}}\left( z \right)|\exp \left( {i{q_z}z} \right)\left| {{\phi _{{{\text{n}}^\prime }}}\left( z \right)} \right\rangle  = {\mathcal{C}_{\text{n}}}{\mathcal{C}_{{{\text{n}}^\prime }}}\int\limits_0^{ + \infty } {{z^{4{\text{s}}}}\exp \left( {i{q_z}z - \dfrac{{{z^2}}}{{\ell _z^2}}} \right)\mathcal{L}_{{{\text{n}}^\prime }}^\alpha \left( {\dfrac{{{z^2}}}{{\ell _z^2}}} \right)\mathcal{L}_{\text{n}}^\alpha \left( {\dfrac{{{z^2}}}{{\ell _z^2}}} \right)dz} .
\end{align}

Next, for simplicity and convenience in numerical calculations, we consider the electron’s transition from ground state to first excited state, i.e. ${\text{n}} = 0,{{\text{n}}^\prime } = 1$. Using the properties of the associated Laguerre function and integrals in \cite{tpb}, we can rewrite Eq. \eqref{j2} as 
\begin{align}
    {\left. {{{\text{J}}_2}} \right|_{{\text{n}} = 0,{{\text{n}}^\prime } = 1}} = \ell _z^{1 + 4{\text{s}}}\Gamma \left( {2{\text{s}} + \dfrac{3}{2}} \right)\left[ {{}_1{F_1}\left( {2{\text{s}} + \dfrac{1}{2};\dfrac{1}{2}; - \dfrac{1}{4}q_z^2\ell _z^2} \right) - {}_1{F_1}\left( {2{\text{s}} + \dfrac{3}{2};\dfrac{1}{2}; - \dfrac{1}{4}q_z^2\ell _z^2} \right)} \right], \label{j22}
\end{align}
where, ${}_1{F_1}\left( {a;b;z} \right)$ is the Kummer confluent hypergeometric function. 

Finally, put \eqref{j22} and \eqref{j1} in \eqref{jnn}, then substitute ${{\text{J}}_{{\text{n}},{{\text{n}}^\prime }}}\left( {\mathbf{q}} \right)$ in Eq. \eqref{gnn} and performing some algebraic transformations, we obtain the expression for the overlap integral over the wave vector of the phonon in Eq. \eqref{gnn} with ${\text{n}} = 0,{{\text{n}}^\prime } = 1$ as follows 
\begin{align}\label{g01}
    {{\text{G}}_{0,1}} = \dfrac{{\pi \left[ {3 + 4\left( {1 + \sqrt {1 + 4{\beta _z}} } \right)} \right]\Gamma \left( {{3 \mathord{\left/
 {\vphantom {3 2}} \right.
 \kern-\nulldelimiterspace} 2} + \sqrt {1 + 4{\beta _z}} } \right)}}{{{2^{{5 \mathord{\left/
 {\vphantom {5 2}} \right.
 \kern-\nulldelimiterspace} 2} + \sqrt {1 + 4{\beta _z}} }}{\ell _z}\Gamma \left( {1 + {{\sqrt {1 + 4{\beta _z}} } \mathord{\left/
 {\vphantom {{\sqrt {1 + 4{\beta _z}} } 2}} \right.
 \kern-\nulldelimiterspace} 2}} \right)\Gamma \left( {2 + {{\sqrt {1 + 4{\beta _z}} } \mathord{\left/
 {\vphantom {{\sqrt {1 + 4{\beta _z}} } 2}} \right.
 \kern-\nulldelimiterspace} 2}} \right)}}.
\end{align}
Conspicuously, the results in Eq. \eqref{g01} when $\beta_z $ goes to zero, we have 
\begin{align}\label{g01p}
    {\left. {{{\text{G}}_{0,1}}} \right|_{{\beta _z} = 0}} = \dfrac{{11}}{{4{\ell _z}}}\sqrt {\dfrac{\pi }{2}} .
\end{align}
This result in Eq. \eqref{g01p} completely coincides with the result obtained in Eq. (19) of Ref. \cite{appa} for the case of a symmetric semi-parabolic quantum well. 

\section{The non-equilibrium distribution functions}\label{hpb}
In this section, we give explicit expressions of the time-independent distribution functions: $\mathcal{F}_{\text{p}}^{(0)}\left( {{\text{n}},{{\mathbf{k}}_ \bot }} \right)$ and $\mathcal{F}_{\text{p}}^{(1)}\left( {{\text{n}},{{\mathbf{k}}_ \bot }} \right)$ in Eqs. \eqref{pb1}, and \eqref{pb2}. In the relaxation time approximation, we rewrite Eq. \eqref{qke} as follows \cite{re5, re6} 
\begin{align}\label{qket}
    \dfrac{{\partial {\mathcal{F}_{{\text{n}},{{\mathbf{k}}_ \bot }}}\left( t \right)}}{{\partial t}}+\left\{ {e{{\mathbf{E}}^{{\text{LPF}}}}\left( t \right) + e{{\mathbf{E}}_0} + \dfrac{{e\hbar }}{{{m_e}c}}\left[ {{{\mathbf{k}}_ \bot } \times {{\mathbf{B}}^{{\text{LPF}}}}\left( t \right)} \right]} \right\}\dfrac{{\partial {\mathcal{F}_{{\text{n}},{{\mathbf{k}}_ \bot }}}\left( t \right)}}{{\hbar \partial {{\mathbf{k}}_ \bot }}} =  - \dfrac{{{\mathcal{F}_{{\text{n}},{{\mathbf{k}}_ \bot }}}\left( t \right) - {\mathcal{F}_0}\left( {{\mathcal{E}_{{\text{n}},{{\mathbf{k}}_ \bot }}}} \right)}}{{\tau \left( \mathcal{E} \right)}}.
\end{align}

In a linear response to the external fields, we can replace $\dfrac{{\partial {\mathcal{F}_{{\text{n}},{{\mathbf{k}}_ \bot }}}\left( t \right)}}{{ \partial {{\mathbf{k}}_ \bot }}}$ in the left-hand side of Eq. \eqref{qket} with $\dfrac{{\partial {\mathcal{F}_0}\left( {{\mathcal{E}_{{\text{n}},{{\mathbf{k}}_ \bot }}}} \right)}}{{ \partial {{\mathbf{k}}_ \bot }}}$ as 
\begin{align}\label{lref}
    \dfrac{{\partial {\mathcal{F}_{{\text{n}},{{\mathbf{k}}_ \bot }}}\left( t \right)}}{{\partial {{\mathbf{k}}_ \bot }}} \to \dfrac{{\partial {\mathcal{F}_0}\left( {{\mathcal{E}_{{\text{n}},{{\mathbf{k}}_ \bot }}}} \right)}}{{\partial {{\mathbf{k}}_ \bot }}} = \dfrac{{\partial {\mathcal{F}_0}\left( {{\mathcal{E}_{{\text{n}},{{\mathbf{k}}_ \bot }}}} \right)}}{{\partial {\mathcal{E}_{{\text{n}},{{\mathbf{k}}_ \bot }}}}}\dfrac{{\partial {\mathcal{E}_{{\text{n}},{{\mathbf{k}}_ \bot }}}}}{{\partial {{\mathbf{k}}_ \bot }}} = \dfrac{{{\hbar ^2}}}{{{m_e}}}\dfrac{{\partial {\mathcal{F}_0}\left( {{\mathcal{E}_{{\text{n}},{{\mathbf{k}}_ \bot }}}} \right)}}{{\partial {\mathcal{E}_{{\text{n}},{{\mathbf{k}}_ \bot }}}}}{{\mathbf{k}}_ \bot }.
\end{align}

Putting \eqref{lref} into \eqref{qket}, noting that $\left[ {{{\mathbf{k}}_ \bot } \times {{\mathbf{B}}^{{\text{LPF}}}}\left( t \right)} \right] \cdot {{\mathbf{k}}_ \bot } = 0$, we simplify Eq. \eqref{qket} as follows 
\begin{align}\label{qkec}
    \dfrac{{\partial {\mathcal{F}_{{\text{n}},{{\mathbf{k}}_ \bot }}}\left( t \right)}}{{\partial t}} + \dfrac{{e\hbar }}{{{m_e}}}\left[ {{{\mathbf{E}}^{{\text{LPF}}}}\left( t \right) + {{\mathbf{E}}_0}} \right]\dfrac{{\partial {\mathcal{F}_0}\left( {{\mathcal{E}_{{\text{n}},{{\mathbf{k}}_ \bot }}}} \right)}}{{\partial {\mathcal{E}_{{\text{n}},{{\mathbf{k}}_ \bot }}}}}{{\mathbf{k}}_ \bot } =  - \dfrac{{{\mathcal{F}_{{\text{n}},{{\mathbf{k}}_ \bot }}}\left( t \right) - {\mathcal{F}_0}\left( {{\mathcal{E}_{{\text{n}},{{\mathbf{k}}_ \bot }}}} \right)}}{{\tau \left( \mathcal{E} \right)}}.
\end{align}

Putting \eqref{elpf}, \eqref{pb1} and \eqref{pb2} into \eqref{qkec}, after some algebraic transformations, we obtain the distribution functions below 

\begin{align}
  \mathcal{F}_{\text{p}}^{(0)}\left( {{\text{n}},{{\mathbf{k}}_ \bot }} \right) &=  - \dfrac{{e\hbar }}{{{m_e}}}\tau \left( {{\mathcal{E} }} \right)\left( {{{\mathbf{k}}_ \bot } \cdot {{\mathbf{E}}_0}} \right)\dfrac{{\partial {\mathcal{F}_0}\left( {{\mathcal{E}_{{\text{n}},{{\mathbf{k}}_ \bot }}}} \right)}}{{\partial {\mathcal{E}_{{\text{n}},{{\mathbf{k}}_ \bot }}}}}, \\
  \mathcal{F}_{\text{p}}^{(1)}\left( {{\text{n}},{{\mathbf{k}}_ \bot }} \right) &=  - \dfrac{{e\hbar }}{{{m_e}}}\dfrac{{\tau \left( {{\mathcal{E} }} \right)}}{{1 - i\omega \tau \left( {{\mathcal{E} }} \right)}}\left[ {{{\mathbf{k}}_ \bot } \cdot {{\mathbf{E}}^{{\text{LPF}}}}\left( t \right)} \right]\dfrac{{\partial {\mathcal{F}_0}\left( {{\mathcal{E}_{{\text{n}},{{\mathbf{k}}_ \bot }}}} \right)}}{{\partial {\mathcal{E}_{{\text{n}},{{\mathbf{k}}_ \bot }}}}}.
\end{align}

\bibliography{main}

\begin{figure}[!htb]
\centering
\subfigure[][Dependence of the REF on photon energy of LRF when the temperature continuously varies from T = 150K to T = 200K. The red and blue regions correspond to the symmetric and asymmetric quantum well models, respectively. \label{3a}]
  {\includegraphics[scale=0.61]{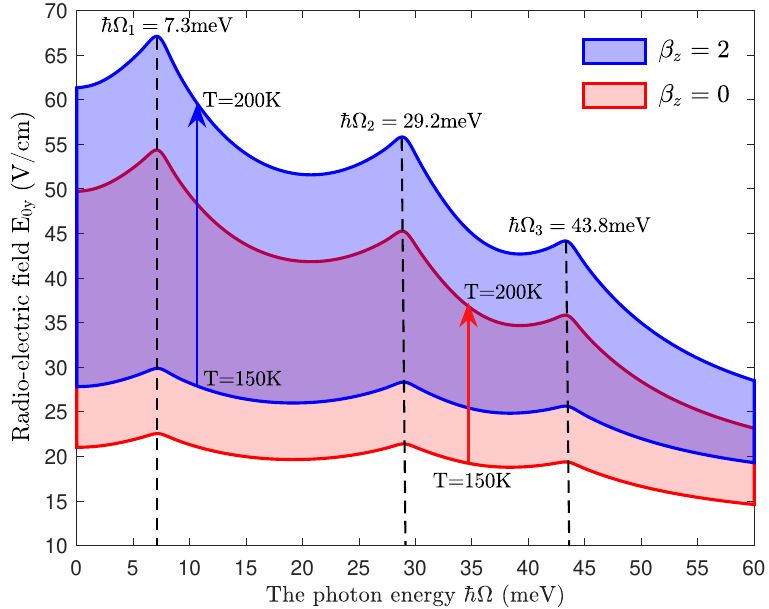}}
\subfigure[][Dependence of the REF on photon energy of LRF at four different values of the intensity of LRF. Here, $\beta_z = 2$, and $T=150$K. \label{3b}]
  {\includegraphics[scale=0.61]{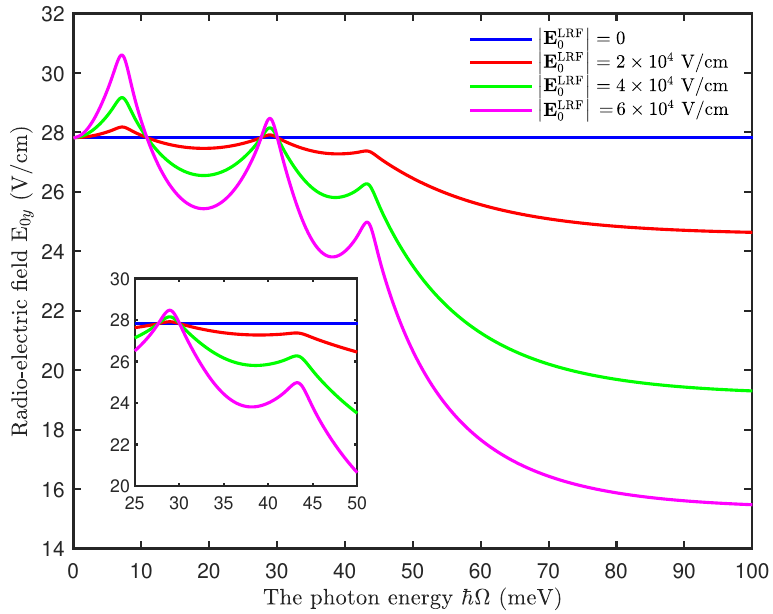}}
\caption{Dependence of REF on LRF photon energy at (a) different temperature regions and (b) different LRF intensity values. Here, ${\omega _z} = 0.1{\omega _0}$.}
\label{fig3}
\end{figure}
\begin{figure}[!htb]
    \centering
    \includegraphics[scale = 0.8]{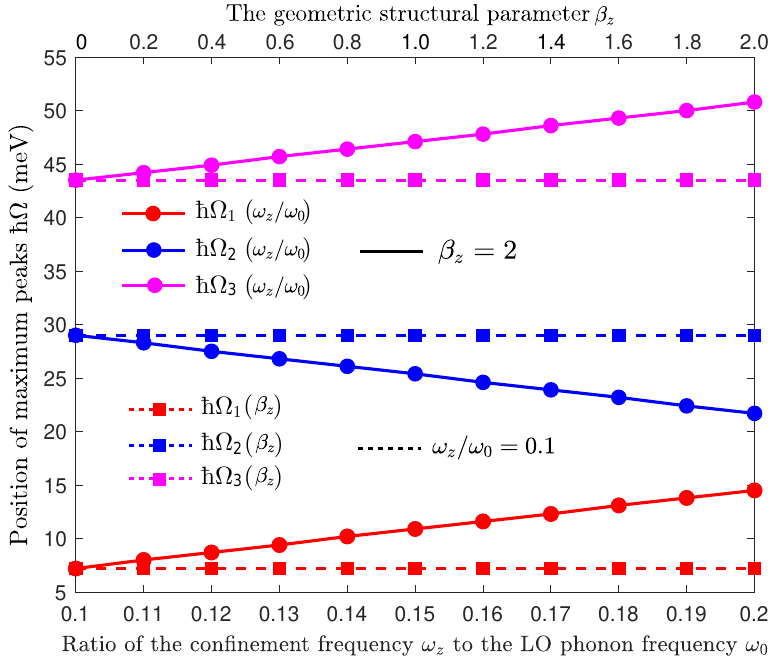}
    \caption{Dependence of the peak positions on the geometric structural parameter $\beta_z$ (solid lines with filled squares) and the confinement frequency $\omega_z$ (dashed lines with filled circles) of an asymmetric semi-parabolic quantum well. Here, $T = 150 \mathrm{K}$, and $\left| {{\mathbf{E}}_0^{{\text{LRF}}}} \right| = 5 \times {10^4}{\text{ V/cm}}$.}  
    \label{fig4}
\end{figure}

\begin{figure}[!htb]
    \centering
    \includegraphics[scale = 0.8]{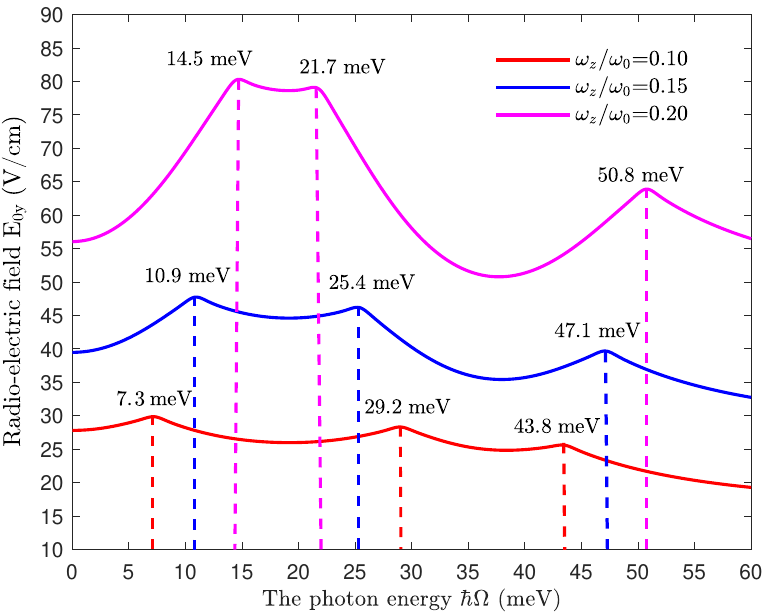}
    \caption{The graph illustrates the results of calculating the position of the maximum peaks shifted when the confinement frequency value changes. Here, $T = 150 \mathrm{K}$, and $\left| {{\mathbf{E}}_0^{{\text{LRF}}}} \right| = 5 \times {10^4}{\text{ V/cm}}$, and $\beta_z = 2$.}  
    \label{fig5}
\end{figure}

\begin{figure}[!htb]
\centering
\subfigure[][ \label{6a}]
  {\includegraphics[scale=0.61]{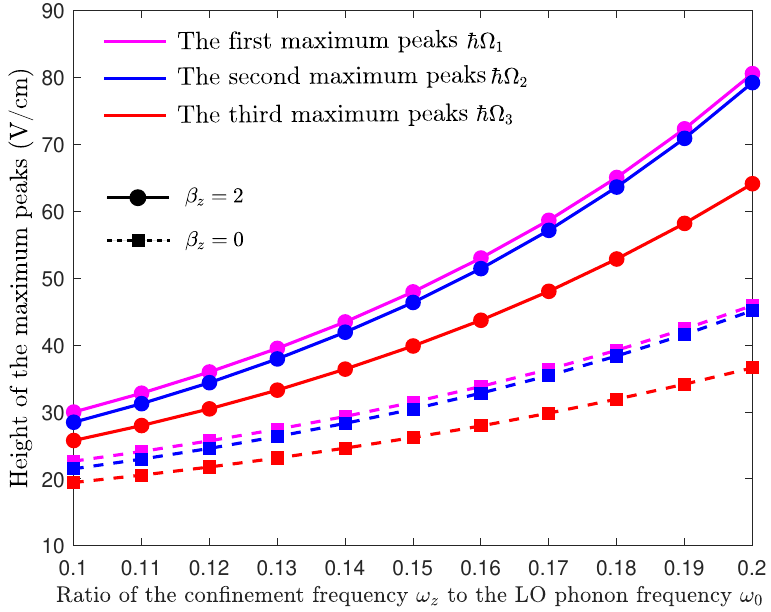}}
\subfigure[][  \label{6b}]
  {\includegraphics[scale=0.61]{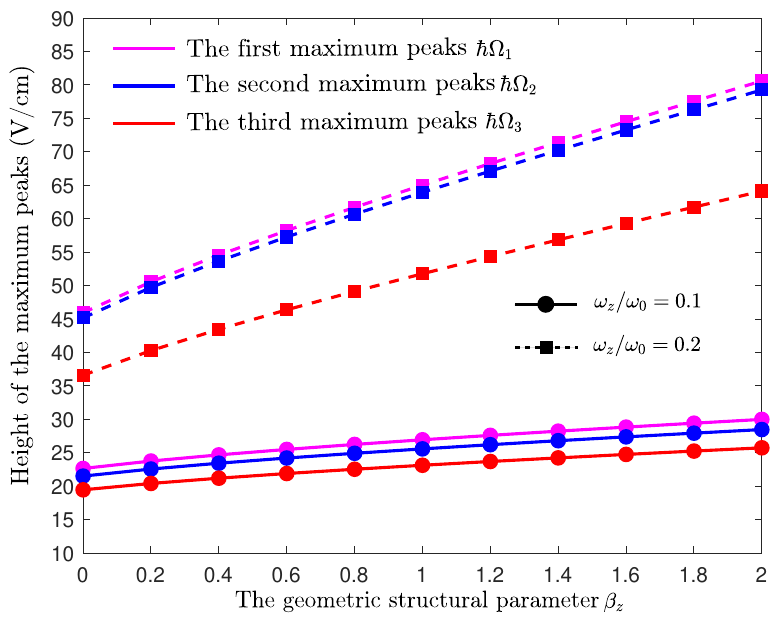}}
\caption{Dependence of the height of the maximum peaks (HMPs) on the parameters of the asymmetric semi-parabolic quantum well. Here, $T = 150 \mathrm{K}$, and $\left| {{\mathbf{E}}_0^{{\text{LRF}}}} \right| = 5 \times {10^4}{\text{ V/cm}}$.}
\label{fig6}
\end{figure}

\begin{figure}[!htb]
\centering
\subfigure[][ \label{7a}]
  {\includegraphics[scale=0.60]{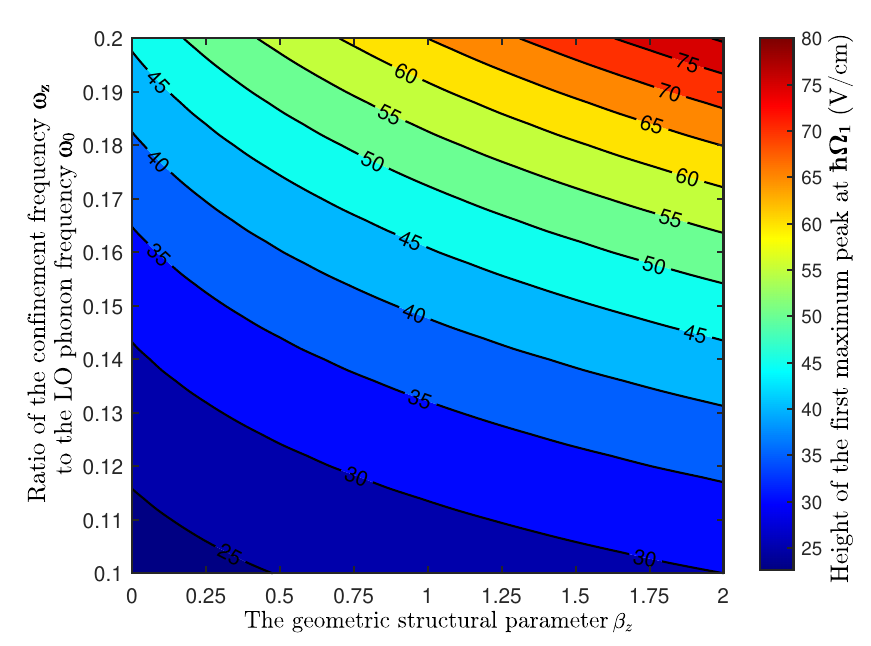}}
\subfigure[][  \label{7b}]
  {\includegraphics[scale=0.60]{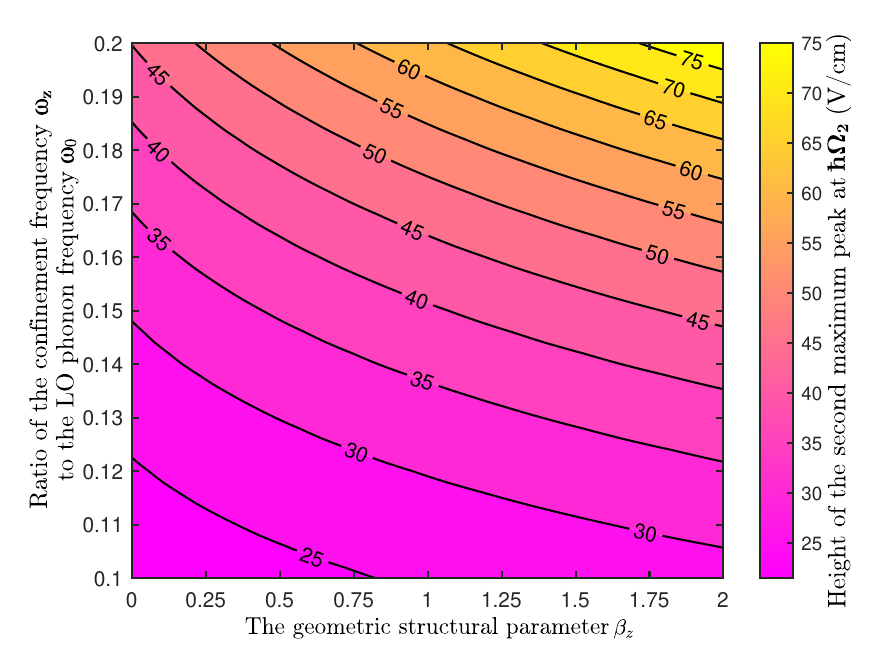}}
  \subfigure[][  \label{7c}]
  {\includegraphics[scale=0.60]{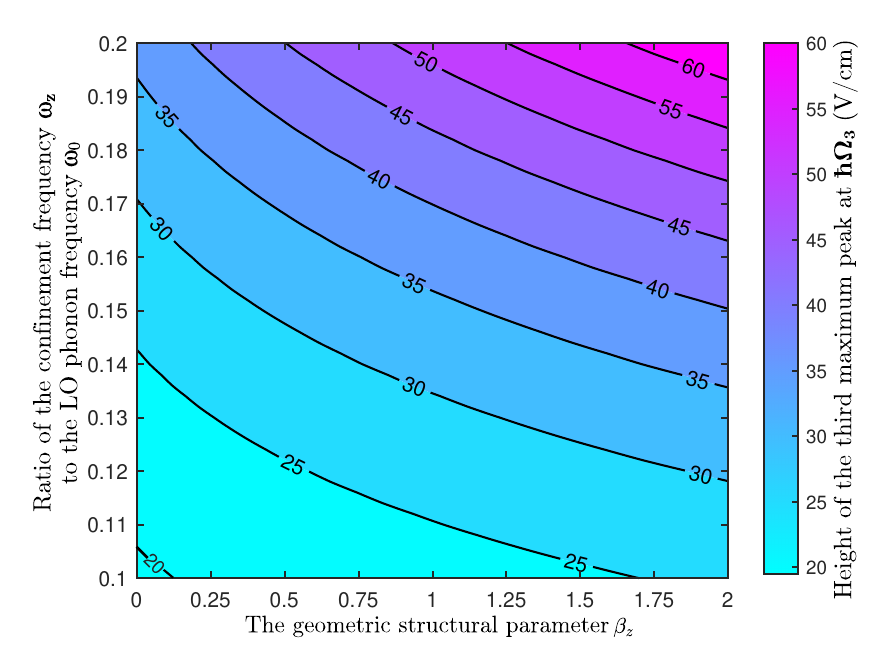}}
    \caption{Density plot of the HMP as a function of the ratio ${{{\omega _z}} \mathord{\left/
 {\vphantom {{{\omega _z}} {{\omega _0}}}} \right.
 \kern-\nulldelimiterspace} {{\omega _0}}}$ and geometric structural parameter $\beta_z$ at the (a) first maximum peak, (b) second maximum peak, and (c) third maximum peak. Here, $T = 150K, \left| {{\mathbf{E}}_0^{{\text{LRF}}}} \right| = 5 \times {10^4}{\text{ V/cm}}$.}
\label{fig7}
\end{figure}

\begin{figure}[!htb]
    \centering
    \includegraphics[scale = 0.8]{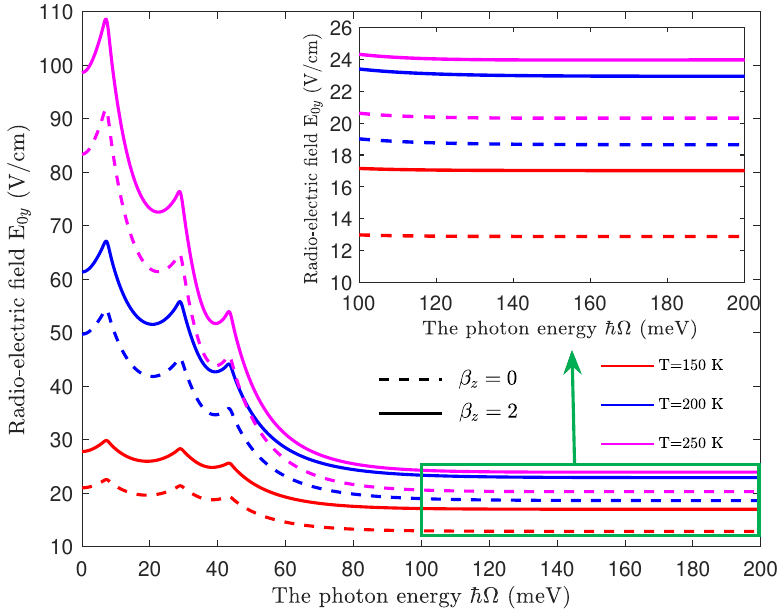}
    \caption{The appearance of the saturated REF value increases with temperature in the photon energy range greater than 100 meV for the cases of symmetric (dashed lines) and asymmetric (solid lines) semi-parabolic quantum wells. Here, ${\omega _z} = 0.1{\omega _0}, \left| {{\mathbf{E}}_0^{{\text{LRF}}}} \right| = 5 \times {10^4}{\text{ V/cm}}$}. 
    \label{fig8}
\end{figure}

\begin{figure}[!htb]
\centering
\subfigure[][Density plot of the saturated REF as a function of the temperature and geometric structural parameter $\beta_z$. Here, ${\omega _z} = 0.1{\omega _0}$. \label{9a}]
  {\includegraphics[scale=0.61]{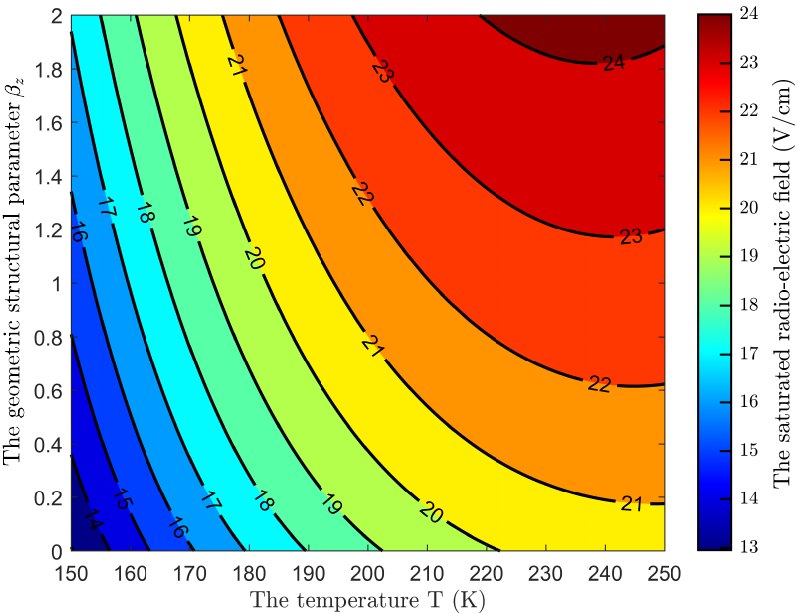}}
\subfigure[][Effect of temperature on saturated REF at different values of confinement frequency and structural parameters. \label{9b}]
  {\includegraphics[scale=0.61]{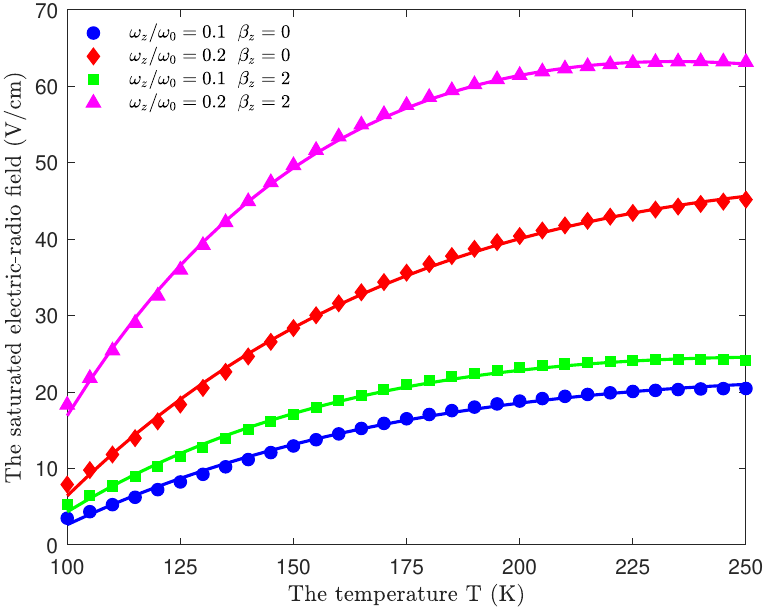}}
\caption{Dependence of temperature on saturated REF. Here, $\left| {{\mathbf{E}}_0^{{\text{LRF}}}} \right| = 5 \times {10^4}{\text{ V/cm}}$. }
\label{fig9}
\end{figure}
\end{document}